\documentclass[preprint,authoryear,sort,a4paper,nopreprintline]{elsarticle}\usepackage[]{graphicx}\usepackage[]{color}
\makeatletter
\def\maxwidth{ %
  \ifdim\Gin@nat@width>\linewidth
    \linewidth
  \else
    \Gin@nat@width
  \fi
}
\makeatother

\usepackage[nodots,compress]{numcompress}
\def\myjournal{Spatial and Spatio-temporal Epidemiology}
\journal{\myjournal}
\def\mytitle{Model-based testing for space-time interaction using point processes:
  An application to psychiatric hospital admissions in an urban area}
\title{%
  \fbox{\vbox{\scriptsize
      This is the accepted manuscript version of the following article:
      Meyer,~S., Warnke,~I., R\"ossler,~W., and Held,~L. (2016).
      Model-based testing for space-time interaction using point processes:
      An application to psychiatric hospital admissions in an urban area.
      \textit{Spatial and Spatio-temporal Epidemiology} \textbf{17}, 15--25,
      which has been published in final form at
      \url{http://dx.doi.org/10.1016/j.sste.2016.03.002}.
    }}\\[1cm]
  \mytitle}

\usepackage{fixltx2e}         
\usepackage[utf8]{inputenc}   
\usepackage[T1]{fontenc}      
\usepackage{lmodern}          

\usepackage{amsmath}          
\usepackage{amsfonts,amssymb} 
\usepackage{mathtools}        
\usepackage{bm}               
\usepackage{bbm}              

\usepackage{subcaption}       
\newcommand{\subfloat}[2][need a sub-caption]{
  \hspace*{\fill}\subcaptionbox{#1}{#2}\hspace*{\fill}}
\usepackage{booktabs}         

\let\code=\texttt
\let\proglang=\textsf
\newcommand{\pkg}[1]{\code{#1}}
\newcommand{\CRANpkg}[1]{\href{http://CRAN.R-project.org/package=#1}{\pkg{#1}}}
\newcommand{\abs}[1]{\lvert#1\rvert}
\newcommand{\norm}[1]{\lVert#1\rVert}

\newcommand{\dif}{\,\mathrm{d}}

\newcommand{\ind}{\mathbbm{1}}

\usepackage[
  pdftitle={\mytitle},
  pdfauthor={Sebastian Meyer},
  pdfsubject={\myjournal},
  colorlinks=true, linkcolor=blue, urlcolor=purple, citecolor=teal
]{hyperref}

\IfFileExists{upquote.sty}{\usepackage{upquote}}{}
\begin{document}


\begin{frontmatter}

\author[EBPI]{Sebastian Meyer\corref{cor1}} 
\cortext[cor1]{Corresponding author
}
\ead{sebastian.meyer@uzh.ch}

\author[PUK1]{Ingeborg Warnke}

\author[PUK2]{Wulf Rössler}

\author[EBPI]{Leonhard Held}
\ead{leonhard.held@uzh.ch}

\address[EBPI]{Epidemiology, Biostatistics and Prevention Institute (EBPI),
  University of Zurich, Hirschengraben 84, 8001 Zürich, Switzerland}
\address[PUK1]{Department of Psychiatry, Psychotherapy and Psychosomatics (DPPP),
  University Hospital of Psychiatry, Lenggstrasse 31, 8032 Zürich, Switzerland}
\address[PUK2]{University Hospital of Psychiatry, Militärstrasse 8, 8021 Zurich, Switzerland}

\begin{abstract}
Spatio-temporal interaction is inherent to cases of infectious diseases and
occurrences of earthquakes,
whereas the spread of other events, such as cancer or crime,
is less evident. Statistical significance tests
of space-time clustering
usually assess the
correlation between the spatial and temporal (transformed) distances
of the events.
Although appealing through simplicity, these classical tests do not adjust for
the underlying population nor can they account for a distance decay of
interaction. We propose to use the framework of an endemic-epidemic point
process model to jointly estimate a background event rate explained by
seasonal and areal characteristics,
as well as a superposed epidemic component representing the hypothesis of interest.
We illustrate this new model-based test for space-time interaction
by analysing psychiatric inpatient admissions in Zurich, Switzerland (2007--2012).
Several socio-economic factors were found to be associated with the admission
rate, but there was no evidence of general clustering of the cases.
\end{abstract}

\begin{keyword}
Spatio-temporal point process\sep
Knox test\sep Mantel test\sep Space-time $K$-function\sep
Global test of clustering\sep
Psychiatric inpatient admissions

\end{keyword}

\end{frontmatter}

\clearpage


\section{Introduction}

Cases of infectious diseases naturally exhibit spatio-temporal interaction.
Once an individual becomes infected, it may cause secondary cases by
transmitting the infectious agent to susceptible individuals.
If the force of infection triggered by an infectious individual decreases with
time and distance, cases are likely to appear in spatio-temporal clusters,
i.e., cases which are close in time tend to also be close in space.
This contagion goes beyond any purely spatial clustering in densely
populated areas or seasonal forcing.
Spatio-temporal clustering can be observed in other fields as well, for example,
seismology \citep{ogata1998},
veterinary epidemiology \citep{ward.carpenter2000},
cancer epidemiology \citep{birch.etal2000,gustafsson.carstensen2000},
criminology \citep{grubesic.mack2008,johnson2010}, 
and transportation research \citep{eckley.curtin2013}.

The spread of various mental health indicators in local social networks has been analysed:
suicide \citep{haw.etal2013},
happiness \citep{fowler.christakis2008},
depression \citep{rosenquist.etal2011},
and even autism \citep{liu.etal2010}.
Usually, this spread occurs within a short time period and in geographically
close living areas.
For hospital admissions of patients with mental disorders,
previous studies have found associations with local socio-economic factors
such as
unemployment rate, 
average income, 
proportion of one-person households,
proportion of foreigners,
or level of urbanisation
\citep[see, e.g.,][]{simone.etal2013}. 
However, to the best of our knowledge, space-time clustering of psychiatric
hospital admissions has not been investigated so far, i.e., if admissions close
in time are also close in space with respect to where the patients live.
The idea is similar to the spread of rumours \citep[Chapter~5]{Daley.Gani1999}, where here the rumour is the news
that someone in the neighbourhood searches professional psychiatric treatment.
For mentally stricken citizens, such news could lower help-seeking barriers,
whereas people with a treatment history might remember own positive experiences
by the knowledge about others' hospital admissions.

The first statistical significance test for ``low intensity epidemicity'' is due
to \citet{knox1963,knox1964}, who investigated space-time interaction of cases
of cleft lip and palate. 
Several variations of his test have since been developed, for example, by generalizing
the test statistic as a measure of correlation of spatial and temporal closeness
\citep{mantel1967}, or by replacing Euclidean distances by $k$~nearest
neighbours \citep{jacquez1996}.
An enhanced version of the Knox statistic derived from point process theory
is the spatio-temporal $K$-function by \citet{diggle.etal1995}.
This approach enables a formal definition of the null hypothesis of no space-time
interaction as a process with independent spatial and temporal components.
However, these classical tests do not account for spatial or temporal
inhomogeneity, e.g., variations in population density or seasonal effects,
and are thus prone to bias
\citep{kulldorff.hjalmars1999,mack.etal2012}.

Second-order characteristics of spatio-temporal point processes, such as the
$K$-function, have recently been generalized to the inhomogeneous case
\citep{gabriel.diggle2009,moeller.ghorbani2012}.
In practice, adjusting the $K$-function for spatial or temporal inhomogeneity
requires a two-step procedure. The crucial first step is to obtain
estimates of the spatial and temporal intensity functions.
As in ``empirical'' point process models
(\citealp[Chapter~12]{Diggle2013}; \citealp{diggle.etal2005}),
non-parametric kernel methods are commonly employed.
Supplied with area-level data on relevant socio-economic factors, we will
instead analyse interaction of events using a \emph{mechanistic}
spatio-temporal point process model. Such models
are especially attractive if covariates can explain part of the
heterogeneity in the observed point pattern,
or if there is an explicit formulation of how past events affect the evolution
of the process. For instance, mechanistic models have recently been applied to a
range of epidemic phenomena, including
residential burglaries \citep{mohler.etal2011},
infectious disease occurrence \citep{meyer.etal2011},
invasive plant species \citep{balderama.etal2012},
and epidemics among dolphins \citep{morris.etal2015}.

We propose to embed the test
for space-time interaction of events in the \emph{endemic-epidemic} point process
regression framework of \citet{meyer.etal2011}, which is implemented in the
open-source \proglang{R}~package \CRANpkg{surveillance} \citep{meyer.etal2014}.
The basic model formulation borrows from both
the Epidemic-Type Aftershock-Sequences (ETAS) model \citep{ogata1998}
and the multivariate time-series model for infectious disease counts proposed by \citet{held.etal2005}.
While the endemic model component reflects background heterogeneity due to,
e.g., population structure and seasonality, the epidemic component makes the
process ``self-exciting'' and causes spatio-temporal interaction.
The basic idea of the proposed model-based
test is to assess the evidence for an epidemic component with a Monte Carlo
permutation approach \citep{besag.diggle1977}.

This paper is organized as follows:
\autoref{sec:classical} briefly reviews the two most popular classical tests
for space-time interaction, the Knox and the Mantel tests,
as well as the space-time $K$-function.
The new test procedure via an endemic-epidemic point process model is
introduced in \autoref{sec:epitest}.
In \autoref{sec:PUK}, we apply the various tests to
psychiatric hospital admissions in the city of Zurich (Switzerland), 2007--2012,
and the supplementary material shows corresponding results for the invasive meningococcal
disease data originally analysed by \citet{meyer.etal2011}.
\autoref{sec:conclusions} concludes the paper and
\ref{sec:software} lists the software used to perform the analyses.

\section{Classical approaches to testing for space-time interaction}
\label{sec:classical}

Given a point pattern $\{(\bm{s}_i, t_i): i = 1,\dotsc,n\}$
with spatial coordinates $\bm{s}_i$ and time points $t_i$
observed in a region $\bm{W}$ during a period $(0,T]$,
classical tests for space-time interaction
basically check if events closer in time also tend to be closer in space.
How closeness is measured varies between the different tests, but many
use the form of test statistic put forward by \citet{mantel1967},
\begin{equation} \label{eqn:classicalT}
  T_0 \propto \sum_{i=1}^n \sum_{j\ne i} a_{ij}^{\text{s}} a_{ij}^{\text{t}} \:,
\end{equation}
where $a_{ij}^{\text{s}}$ and $a_{ij}^{\text{t}}$ are measures of the spatial
and temporal adjacency of the events $i$ and $j$, respectively
\citep{marshall1991}.
These measures are often defined as a function of the respective Euclidean
distances $d_{ij}^{\text{s}} = \norm{\bm{s}_i - \bm{s}_j}$ and
$d_{ij}^{\text{t}} = \abs{t_i - t_j}$.
Larger values of $T_0$ support the
(one-sided) alternative hypothesis of positive association between spatial and
temporal adjacency.

The distribution of $T_0$ under the null hypothesis of no space-time interaction
is determined by a Monte Carlo permutation approach \citep{besag.diggle1977}.
In each of $B = 999$, say, replications, the test statistic is computed
for a random permutation of the time labels while holding the locations fixed.
This destroys any systematic space-time interaction in the data but leaves both
the marginal spatial and temporal distributions unchanged.
The (one-sided) $p$-value for a positive association between spatial and temporal adjacency
(i.e., clustering) is then obtained as the proportion of test statistics greater
than or equal to the observed one. The smallest attainable $p$-value from this
Monte Carlo procedure thus is $1/(B+1) = 0.001$ (the ``resolution'').

In the following subsections, we outline the various choices of adjacency
measures employed by three different tests.
For a broader overview of space-time interaction tests and references to
applications we point to \citet[Chapter~7]{Tango2010} and
\citet[Section~2]{mack.etal2012}.

\subsection{Knox test}

For the Knox test, critical distances in space ($\delta$) and time ($\tau$) have
to be specified to yield a categorization of the distances into ``close''
vs.\ ``not close''. The test statistic is then defined as the number of event
pairs close both in space and time according to these distance thresholds:
\begin{equation} \label{eqn:Knox}
  T_{\text{Knox}} = \frac{1}{2} \sum_{i=1}^n \sum_{j\ne i} \ind(d_{ij}^{\text{s}} \le \delta)
  \ind(d_{ij}^{\text{t}} \le \tau) \:.
\end{equation}
If the point pattern exhibits clustering at the predefined spatio-temporal
scales, the observed number of close pairs will be larger than the expected
number under the null hypothesis of no space-time interaction.

The Knox test is appealing through simplicity but criticized for the
subjectivity in specifying $\delta$ and $\tau$. It is thus often
applied over a range of critical distances \citep{grubesic.mack2008}.

\subsection{Mantel test}

\citet{mantel1967} elaborates on the general use of the test statistic
\eqref{eqn:classicalT} with any suitable distance measures for the application
at hand. This includes the indicator functions of the Knox test,
raw or transformed Euclidean distances such as $1/(d_{ij}+1)$ (to collapse
the range of large distances), but also other distance measures like travel time.
An often used standardized version of Mantel's test statistic is the Pearson
correlation between the spatial and temporal distances of all event pairs, i.e.,
\begin{equation} \label{eqn:Mantel}
  T_{\text{Mantel}} = \frac{1}{n(n-1)-2} \sum_{i=1}^n \sum_{j\ne i}
  \frac{d_{ij}^{\text{s}} - \bar{d}_{\text{s}}}{\hat\sigma_{d^{\text{s}}}}
  \frac{d_{ij}^{\text{t}} - \bar{d}_{\text{t}}}{\hat\sigma_{d^{\text{t}}}} \:,
\end{equation}
where the $\bar{d}_{\cdot}$ and $\hat\sigma_{d^{\cdot}}$ symbols denote the respective sample
means and standard deviations of the $n(n-1)/2$ pairwise distances.
Note that the standardized Mantel test statistic formulated in
\citet[Section~1.2]{jacquez1996}, \citet[Section~2.1]{ward.carpenter2000}, and
\citet[Section~2.2]{mack.etal2012} does \emph{not} exactly correspond to the
Pearson correlation \eqref{eqn:Mantel} as used in this paper and in many open-source
software packages implementing the Mantel test.\footnote{
  We have inspected the \proglang{R} packages
  \CRANpkg{ecodist}, \CRANpkg{vegan}, and \CRANpkg{ade4} 
  (all available at \url{http://CRAN.R-project.org/}),
  as well as the \proglang{python} packages
  \pkg{PySAL} (\url{http://pysal.org/})
  and \pkg{scikit-bio} (\url{https://github.com/biocore/scikit-bio/}). 
}

Unlike the Knox test, the standardized Mantel test does not require the
specification of distance thresholds. However, it assesses a linear
relationship between spatial and temporal 
distances, which might not be appropriate over the whole distance range
(the clustering is expected to occur towards the
origin in a plot of spatial against temporal distance). The reciprocal
transformation of the pairwise distances suggested by \citet{mantel1967} would
then again depend on the choice of a suitable constant to avoid division by zero.

\subsection{Space-time $K$-function analysis}

The space-time $K$-function $K(\delta,\tau)$ essentially interprets the Knox
statistic \eqref{eqn:Knox} as a function of the critical distances $\delta$ and
$\tau$ while accounting for edge effects \citep{diggle.etal1995}.
For stationary point processes, the $K$-function is proportional to the
expected number of further events occurring within distance $\delta$ and time
$\tau$ of an arbitrary event \citep[Chapters~10 and~11]{Diggle2013}.
An approximately unbiased estimator of the $K$-function is
\begin{equation} \label{eqn:stK}
  \hat{K}(\delta,\tau) = \frac{\abs{W}T}{n(n-1)} \sum_{i=1}^n \sum_{j\ne i} w_{ij} v_{ij}
  \ind(d_{ij}^{\text{s}} \le \delta) \ind(d_{ij}^{\text{t}} \le \tau) \:,
\end{equation}
where $w_{ij}$ and $v_{ij}$ are edge-correction weights \citep[Section~3]{diggle.etal1995}.
Note that this estimator is proportional to the Knox statistic \eqref{eqn:Knox}
if the weights all equal unity. Similar estimators $\hat{K}_s(\delta)$ and
$\hat{K}_t(\tau)$ exist for the purely spatial and temporal component processes,
respectively. The test statistic is then derived from the property that the
spatio-temporal $K$-function factorizes into the component $K$-functions under
the null hypothesis of no space-time interaction. Specifically,
\begin{equation} \label{eqn:Dhat}
  \hat{D}(\delta,\tau) = \hat{K}(\delta,\tau) - \hat{K}_s(\delta) \, \hat{K}_t(\tau)
\end{equation}
has an expectation of zero for any given thresholds $\delta$ and $\tau$.
\citet{diggle.etal1995} recommend a perspective plot of the $\hat{D}$ surface
to gain more insight into the relevant scales of
spatio-temporal interaction than the previous tests.
An omnibus test over a set of spatial and temporal scales is obtained by
\begin{equation} \label{eqn:stKtest}
  T_{\text{Diggle}} = \sum_{\delta} \sum_{\tau} \hat{D}(\delta, \tau)
\end{equation}
and evaluating statistical significance by the common Monte Carlo permutation
approach.

This test is used far less frequently than the Knox or Mantel tests, since a
(polygonal) representation of the observation region $\bm{W}$ is required in 
addition to the event times and locations to compute the edge-correction
weights, which also needs specialized software.
Some studies such as \citet{mcnally.etal2006} have used a
simplification of this test procedure towards an omnibus Knox test by ignoring
the edge-correction weights. 

\section{Model-based assessment of space-time interaction}
\label{sec:epitest}

The classical Knox and Mantel tests have the advantage of working solely on the
observed event times and locations without the need for additional data.
In practice, however, the question of space-time interaction is often
accompanied by questions of how the event rate varies over time or space, how
far the events interact, or if specific events trigger a higher amount of
clustering.
To address these issues, we propose to embed the test for
space-time interaction in a regression framework for spatio-temporal point
patterns of epidemic phenomena
\citep{meyer.etal2011,meyer.held2013}.

\subsection{Spatio-temporal point process model}
\label{sec:twinstim}

The considered spatio-temporal point process model describes the
conditional intensity $\lambda(\bm{s},t)$ for an event at location $\bm{s}$ and
time $t$ (given the history of the process) by a superposition of an endemic and
an epidemic component:
\begin{align} \label{eqn:twinstim}
  \lambda(\bm{s},t) &= \rho_{[\bm{s}][t]} \, \nu_{[\bm{s}][t]} +
  \sum_{j \in I(\bm{s},t)} \eta_j \, f(\norm{\bm{s}-\bm{s}_j}) \, g(t-t_j) \:,\\
  I(\bm{s},t) &= \big\{ j : t_j < t \:\wedge\: t-t_j \le \tau_j
  \:\wedge\: \norm{\bm{s}-\bm{s}_j} \le \delta_j \big\} \:.\nonumber
\end{align}
The endemic component reflects the background rate of new events (cases) as explained by
the offset $\rho_{[\bm{s}][t]}$, usually population density, and effects of
other local and/or time-varying characteristics in the log-linear predictor
$\nu_{[\bm{s}][t]}$. Here the spatial and temporal indices $[\bm{s}][t]$ refer to the
regions and periods across which the covariates are collected, e.g., a district
$\times$ week grid. Interpretation of the stand-alone endemic model component is
thus straightforward since it is equivalent to a Poisson regression model for
the aggregated number of events across the cells of the chosen grid
\citep[Section~3.1]{meyer.etal2014}. Note that this data-driven formulation
leads to a piecewise constant endemic intensity, which -- depending on the chosen
grid and the underlying event-generating process -- may not be a
maintainable simplification.

The second, observation-driven epidemic component adds ``infection pressure''
from the set~$I(\bm{s},t)$ of past events and thus causes
spatio-temporal interaction. During its infectious period of length~$\tau_j$
and within its spatial interaction radius~$\delta_j$,
the model assumes each event~$j$ to trigger secondary cases at a rate
proportional to a predictor~$\eta_j$ of event marks~$\bm{m}_j$.
The typical decay of infection pressure with increasing spatial and temporal
distance from the infective event is modelled by parametric interaction
functions~$f$ and~$g$, respectively \citep[Section~4]{lawson.leimich2000}.
Depending on the application at hand, these could simply be assumed constant,
or, e.g., a power-law distance decay for~$f$ could be chosen to reflect human
travel behaviour on larger scales \citep{meyer.held2013}.
The (possibly infinite) upper bounds~$\tau_j$ and~$\delta_j$ provide a way of
modelling event-specific interaction ranges, but since these have to be
specified a common assumption is $\tau_j \equiv \tau$ and
$\delta_j \equiv \delta$.

The spatio-temporal point process model \eqref{eqn:twinstim}
corresponds to a branching process with immigration,
where part of the event rate is due to the endemic (immigration) component
reflecting sporadic cases caused by unobserved sources of infection.
The expected number of offspring an event generates according to the
``triggering function'' $\eta_j \, f(\norm{\bm{s}-\bm{s}_j}) \, g(t-t_j)$
can thus be interpreted as a model-based effective reproduction number.
This number is obtained as the integral of the triggering function over
the observed interaction period $(t_j, t_j+\tau_j] \cap (0,T]$ and region
$b(\bm{s}_j, \delta_j) \cap \bm{W}$, where $b(\bm{s}_j, \delta_j)$ denotes the
disc centered at $\bm{s}_j$ with radius $\delta_j$.

\subsection{Testing for epidemic behaviour}

A standard option to assess if the process at hand shows epidemic
behaviour is a likelihood ratio test between the full model and the
corresponding endemic-only model. The associated test statistic is
$D = -2 \log( L_{\text{endemic}} / L_{\text{full}} )$,
where $L_{\text{endemic}}$ is the maximized likelihood of
the endemic-only model given the observed point pattern,
and $L_{\text{full}}$ the corresponding likelihood of the
full model with an epidemic component.
Originally, \citet{meyer.etal2011} used a log-link for the epidemic predictor,
i.e., $\eta_j = \exp(\gamma_0 + \bm{\gamma}^\top \bm{m}_j)$, 
in which case the null hypothesis of an endemic-only model with $\eta_j = 0$ is
on the boundary of the parameter space \citep{self.liang1987}.
Using the identity link for $\eta_j$ avoids this problem and allows for the
model to represent an inhibition process with a negative epidemic intensity
(as long as $\lambda(\bm{s},t) \ge 0$ at all $(\bm{s},t)$ and
$\lambda(\bm{s}_i,t_i) > 0$ at all events $i=1,\dotsc,n$).

However, the likelihood ratio test is still non-standard since parameters of
the interaction functions~$f$ and~$g$ are not identifiable under the null
hypothesis $(\gamma_0,\bm{\gamma}) = \bm{0}$. 
More importantly, we will see that including an epidemic component can
improve the fit for reasons other than pure presence of space-time interaction
--- if the data provide evidence against the piecewise constant endemic intensity.
For these reasons, we will determine the null distribution of the test statistic
by a Monte Carlo permutation approach just as for the classical significance
tests described in Section~\ref{sec:classical}.
Note that this is only valid for models with a separable endemic intensity,
i.e., independent spatial and temporal background processes.
For each of $B$ random relabelings of the event times, the full model and the
endemic-only model have to be re-estimated to determine the test statistic for
the permuted point pattern. We perform the test conditionally on the estimated
interaction functions $f$ and $g$ to avoid identifiability issues for permuted
data with a naturally low epidemic intensity, and to drastically reduce the
computational burden (the point process likelihood requires the integration of
$f(\norm{\bm{s} - \bm{s}_j})$ over $\bm{W} \cap b(\bm{s}_j, \delta_j)$,
see \citealp[Section~3.1]{meyer.etal2014}).

An advantage of the permutational approach
is that we can replace the likelihood ratio statistic $D$ with a more
accessible quantity, for example the reproduction number described above.
Since the expected number of offspring implied by the point process model
\eqref{eqn:twinstim} is event-specific, we use
\begin{equation} \label{eqn:simpleR0}
  T_R = \hat{\gamma}_0
  \left[ \int_{b(\bm{0}, \delta)} \hat{f}(\bm{s}) \dif\bm{s} \right]
  \left[ \int_0^\tau \hat{g}(t) \dif t \right]
\end{equation}
as a basic effect size,
i.e., the estimated number of secondary cases triggered by an event with
$\bm{m}_j = \bm{0}$ during an infectious
period of length $\tau$ within a surrounding region of radius $\delta$.

\section{Psychiatric hospital admissions in Zurich, 2007--2012}
\label{sec:PUK}

We now investigate contagion of psychiatric hospital admissions in an urban
catchment area using the tests for space-time interaction described in
Sections~\ref{sec:classical} and~\ref{sec:epitest}.
The spread of the news about a psychiatric hospital admission is thereby assumed
to be most likely in close neighbourhood of the patient's residence
within a short period of time. Specifically, we assume a maximum interaction
radius of $\delta = 250$ metres and an infectious period
of $\tau = 14$ days starting from admission.
For the main hypothesis, we make no distinction with respect to the patients'
diagnoses. However, since contagion might be restricted to a
specific subset, we also apply the tests to admissions of the
two main diagnostic subgroups separately:
schizophrenia (ICD-10 F2x) and affective disorders (F3x).

The following section describes the spatio-temporal point pattern of hospital
admissions and the additional socio-economic data to be adjusted for in the
point process model. Subsequently, the classical and model-based tests for
space-time interaction are applied, followed by a discussion of the results.

\subsection{Materials}

\subsubsection{Case reports}

We use data of inpatient episodes from the central register of the university
hospital of psychiatry (PUK) in Zurich, Switzerland, from the years 2007 to 2012.
The catchment area of the PUK includes about 450\,000 to 500\,000 inhabitants.
The PUK is one of six psychiatric institutions which serve a defined catchment
area in the canton of Zurich and which treat the whole spectrum of mental health
problems. The PUK covers almost 40\% of the treatment episodes of these
institutions.


\begin{figure}

{\centering \subfloat[Administrative quarter boundaries of Zurich and their populated area shaded according to the number of patients per 1\,000 inhabitants. The black triangle marks the location of the PUK and the thick boundary line encloses inner urban quarters (see text).\label{fig:pukepi1}]{\includegraphics[width=0.47\linewidth]{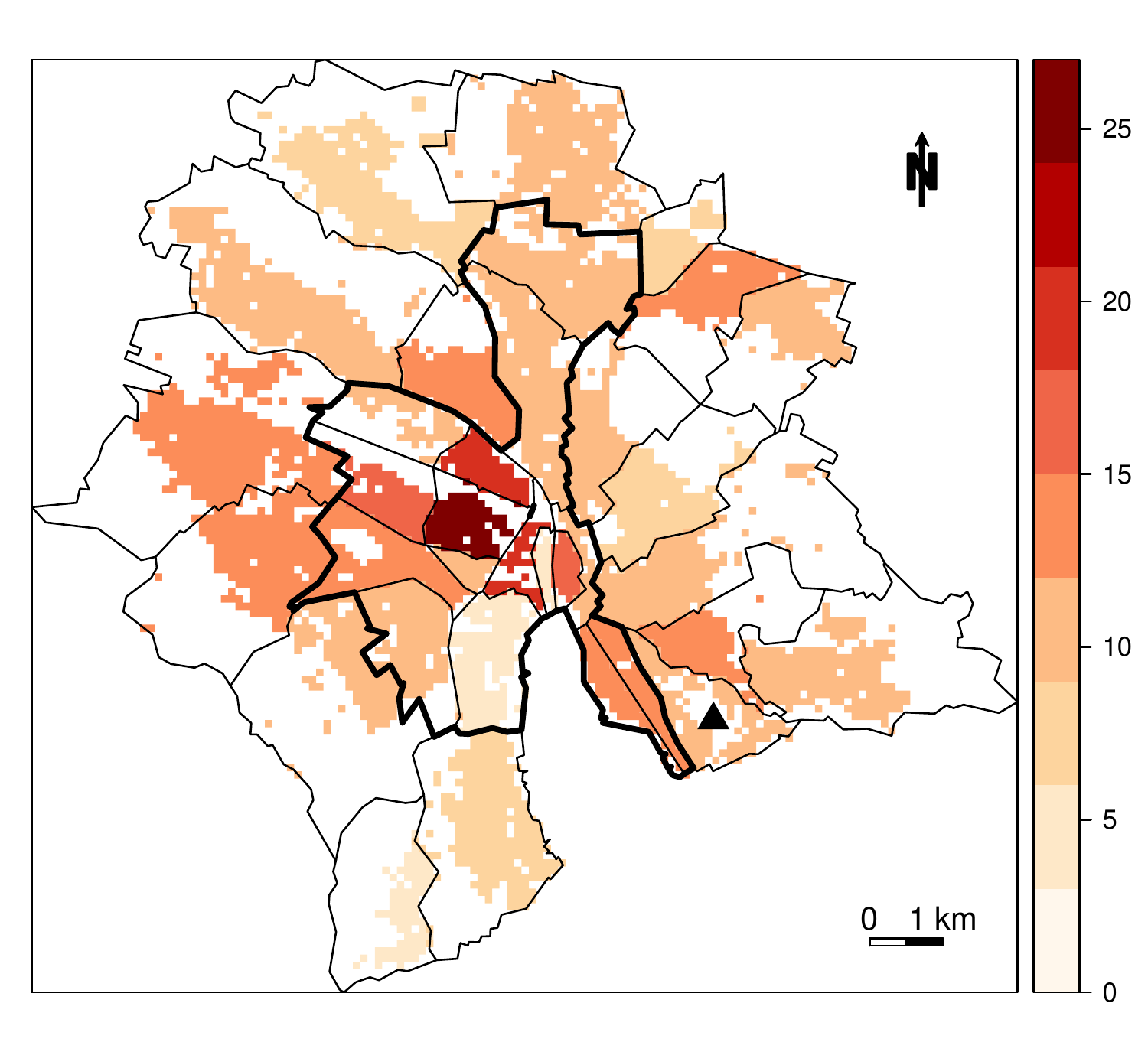} }\subfloat[Monthly and cumulative number of admissions stratified by diagnosis group.\label{fig:pukepi2}]{\includegraphics[width=0.47\linewidth]{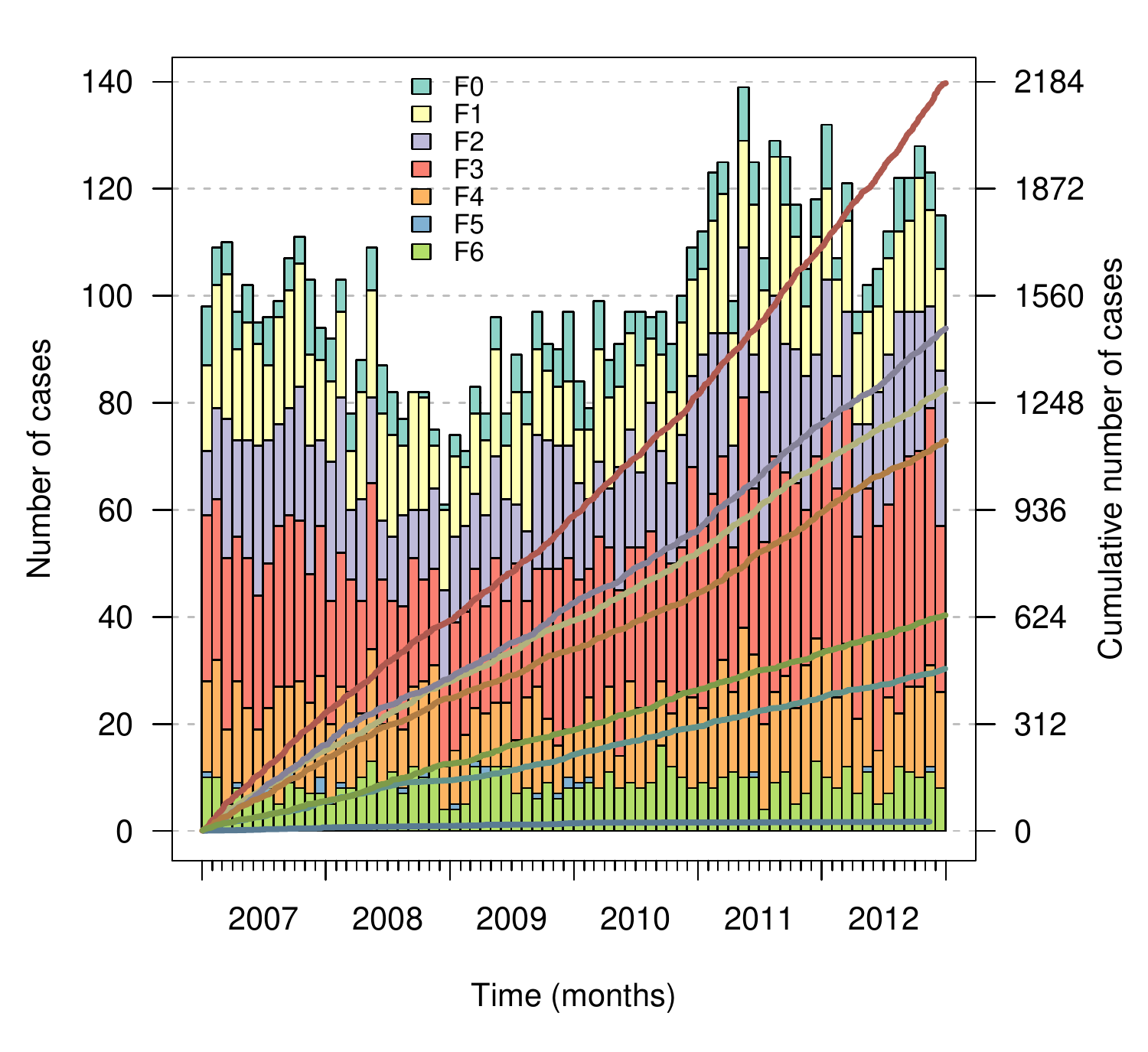} }

}

\caption{Spatio-temporal characteristics of the 7\,202 admissions to the PUK, 2007--2012.}\label{fig:pukepi}
\end{figure}

In total the PUK has registered 23\,290 admissions of
12\,280 different patients
during the years 2007--2012.
The patient's entry date defines the event time $t_i$, and the geographic coordinates
of the patient's residence in metre units define the corresponding event location $\bm{s}_i$.
We include all voluntary admissions of adult patients living in the city of
Zurich at the time of admission, where the primary diagnosis (as qualified at
discharge) is in the range F0x--F6x. Of these
10\,320
records, we exclude all admissions without proper coordinate information
(e.g., unknown/ambiguous address, homeless patients),
and all re-admissions within 14 days (to not infer self-infections).
The remaining sample to analyse consists of 7\,202 admissions
of 4\,498 patients.
Figure~\ref{fig:pukepi} shows the quarter-wise number of patients per
1\,000 inhabitants as well as the monthly number of admissions.\footnote{
  Ethical approval for this study was obtained from the
  ethics committee of the Canton of Zurich,
  provided that the data will only be published in adequately aggregated form.
}
The F2 and F3 subsets contain
1\,465 (856) and
2\,180 (1\,602)
admissions (patients), respectively.

\subsubsection{Population}

The Swiss Federal Statistical Office provided us with population data of the
city of Zurich on a $100m \times 100m$ (i.e., 1 hectare) grid with a total of
392\,435 inhabitants.
With these data, the observation region $\bm{W}$ of the point
process can be defined as the populated area of Zurich
(the shaded area in Figure~\ref{fig:pukepi1}).
Using the population-based observation region is more adequate than
administrative boundaries, which also encompass the large unpopulated areas
of the surrounding hills where no events can occur.

\subsubsection{Quarter-specific socio-economic data}

Socio-economic characteristics of Zurich's 34 quarters
were provided by the Statistics Office of the city of Zurich.
Some of the variables were available on a yearly basis but
there was not much variation across the years. We thus consistently
use the data of the year 2010, which is also the year where most of the
time-constant data has been collected as part of that year's census.
The following socio-economic characteristics are used:
proportion of foreigners,
proportion aged 40 to 64 years,
proportion of one-person households,
employment rate (proportion employed among the population aged 15 years and over),
proportion of inhabitants with minimum (obligatory) education,
and mean yearly income per taxpayer.
Summary statistics for these quarter-level variables are given in
Table~\ref{tab:endemicdata_2010}.
An extra indicator variable was constructed from land use statistics to
distinguish between urban and rural quarters:
each of the inner urban quarters, i.e., those surrounded by a
thick line in Figure~\ref{fig:pukepi1}, has more than 40\% of its area
occupied by buildings and infrastructure and less than 2\% forest area.

\begin{table}[ht]
\centering
\caption{Summary statistics for socio-economic variables of Zurich's 34 quarters.} 
\label{tab:endemicdata_2010}
\begingroup\footnotesize
\begin{tabular}{rrrrrrr}
  \toprule
 & Min. & 1st Qu. & Median & Mean & 3rd Qu. & Max. \\ 
  \midrule
\% Foreigners & 20.1 & 26.0 & 29.4 & 29.9 & 34.4 & 41.0 \\ 
  \% 40--64 years & 26.0 & 29.4 & 31.4 & 31.8 & 33.8 & 41.4 \\ 
  \% One-person households & 12.1 & 20.2 & 23.4 & 23.8 & 26.4 & 39.1 \\ 
  Employment rate & 54.6 & 62.4 & 66.1 & 66.8 & 71.2 & 83.2 \\ 
  \% Low-level education & 14.0 & 19.9 & 31.8 & 29.2 & 38.3 & 47.0 \\ 
  Mean income (in 1000 CHF) & 35.0 & 40.8 & 46.2 & 47.1 & 54.0 & 69.5 \\ 
   \bottomrule
\end{tabular}
\endgroup
\end{table}

\subsubsection{Density of psychiatrists by city district}

The number of psychiatric practices by city district (decomposing Zurich into 12
larger subregions) was obtained from the medical society of the canton of Zurich.
To make use of this information when modelling the endemic risk at the smaller quarter
level, the number was transformed to a density per hectare of populated district
area. This density is then assumed to hold for all quarters of the corresponding
district. 
It ranges from
0.003 to 0.437
practices per hectare.
Aggregated over the whole populated area of Zurich, there are 
0.08
psychiatric practices per hectare.

\subsection{Classical significance tests}

Table~\ref{tab:knoxtab_PUK} shows the contingency tables underlying the
conducted Knox tests. The associated $p$-values are
0.79 (all cases),
0.93 (F2), and
0.81 (F3), respectively,
suggesting no evidence for space-time interaction at the predefined scales,
neither in the overall sample nor within the main diagnoses subgroups.

\begin{table}[ht]
\centering
\caption{Contingency tables of the spatial and temporal distances of all pairs
  of PUK admissions, and in the subgroups of schizophrenic and affective
  disorders, respectively.
  Each table's caption gives the expected number $E$ of close pairs in the
  absence of space-time interaction and the permutation-based $p$-value.}
\label{tab:knoxtab_PUK}
\subfloat[All cases ($E = 2448$,
$p = 0.79$).]{
\begingroup\footnotesize
\begin{tabular}{r|rr|r}
  \toprule
 & \multicolumn{2}{|c|}{km apart} & \\
days apart & $\le$ 0.25 &  $>$ 0.25 & $\sum$ \\ 
  \midrule
$\le$ 14 & 2\,408 & 346\,953 & 349\,361 \\ 
   $>$ 14 & 179\,271 & 25\,402\,169 & 25\,581\,440 \\ 
   \midrule
$\sum$ & 181\,679 & 25\,749\,122 & 25\,930\,801 \\ 
   \bottomrule
\end{tabular}
\endgroup

}

\medskip
\subfloat[F2-cases ($E = 134$,
$p = 0.93$).]{
\begingroup\footnotesize
\begin{tabular}{r|rr|r}
  \toprule
 & \multicolumn{2}{|c|}{km apart} & \\
days apart & $\le$ 0.25 &  $>$ 0.25 & $\sum$ \\ 
  \midrule
$\le$ 14 & 119 & 14\,388 & 14\,507 \\ 
   $>$ 14 & 9\,774 & 1\,048\,099 & 1\,057\,873 \\ 
   \midrule
$\sum$ & 9\,893 & 1\,062\,487 & 1\,072\,380 \\ 
   \bottomrule
\end{tabular}
\endgroup

}
\subfloat[F3-cases ($E = 226$,
$p = 0.81$).]{
\begingroup\footnotesize
\begin{tabular}{r|rr|r}
  \toprule
 & \multicolumn{2}{|c|}{km apart} & \\
days apart & $\le$ 0.25 &  $>$ 0.25 & $\sum$ \\ 
  \midrule
$\le$ 14 & 214 & 32\,564 & 32\,778 \\ 
   $>$ 14 & 16\,156 & 2\,326\,176 & 2\,342\,332 \\ 
   \midrule
$\sum$ & 16\,370 & 2\,358\,740 & 2\,375\,110 \\ 
   \bottomrule
\end{tabular}
\endgroup

}
\end{table}

The standardized Mantel tests yield $p$-values (Pearson correlations given in parentheses) of 
0.092
($r_{\text{all}} = 0.0044$),
0.13
($r_{\text{F2}} = 0.0079$),
and 0.88
($r_{\text{F3}} = -0.0069$),
respectively.
The correlations between spatial and temporal distances are generally very small and
even negative for the subgroup of affective disorders.
Hence, the Mantel test also yields no evidence for space-time interaction.

\begin{figure}[bth]

{\centering \subfloat[All cases ($p =0.94$).\label{fig:stKplot_PUK1}]{\includegraphics[width=0.31\linewidth]{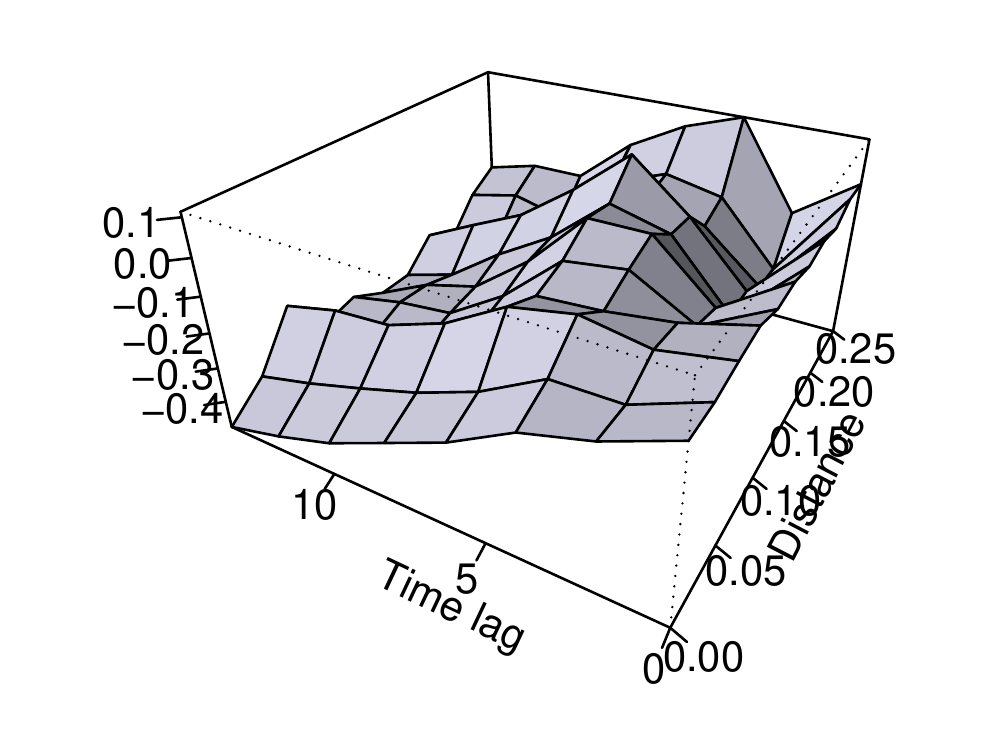} }\subfloat[F2 cases ($p =0.98$).\label{fig:stKplot_PUK2}]{\includegraphics[width=0.31\linewidth]{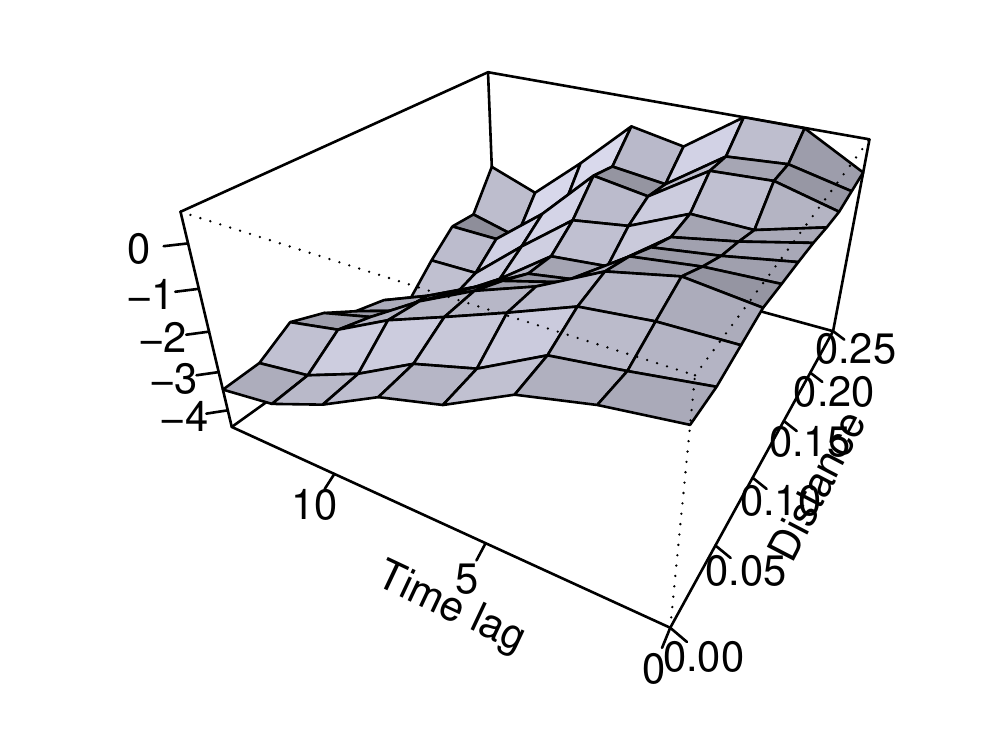} }\subfloat[F3 cases ($p =0.99$).\label{fig:stKplot_PUK3}]{\includegraphics[width=0.31\linewidth]{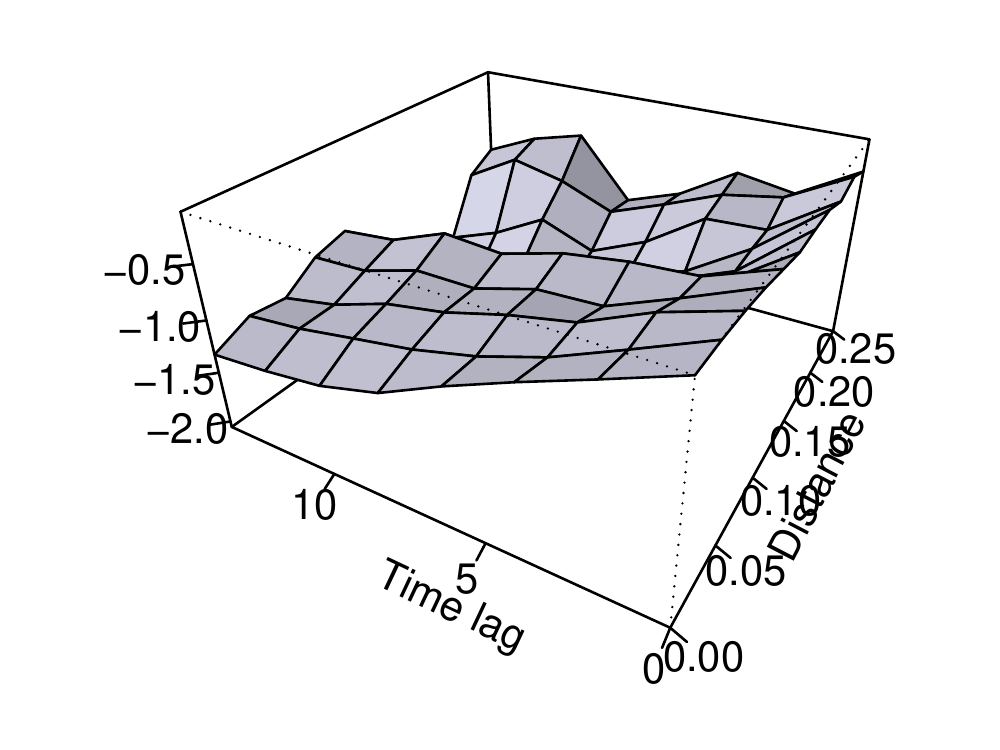} }

}

\caption{Perspective plots of $\hat{D}(\delta,\tau)$ with the permutation-based $p$-value given in the caption.}\label{fig:stKplot_PUK}
\end{figure}

Figure~\ref{fig:stKplot_PUK} shows perspective plots of $\hat{D}(\delta,\tau)$
from Equation~\eqref{eqn:Dhat},
evaluated on an equidistant grid covering the distance range
0 to 250 metres
and the time lags 0 to 14 days,
respectively.
For an infectious process one would expect an excess number of ``close'' cases
compared to an interaction-free process. However, this excess as measured by 
$\hat{D}(\delta,\tau)$ is negative for most interaction ranges and generally
decreases with the time lag $\tau$.
The omnibus test \eqref{eqn:stKtest} based on 999 random
permutations of the event times yields $p$-values of
0.94 (all cases),
0.98 (F2 cases), and
0.99 (F3 cases), respectively,
all indicating absence of evidence for space-time interaction.

\subsection{Endemic-epidemic point process model}

To simultaneously estimate effects of local socio-economic characteristics on
the admission rate and investigate additional space-time interaction of the cases,
we formulate the following endemic-epidemic point process model:
\begin{equation} \label{eqn:pukmodel}
  \lambda(\bm{s},t) =
  \rho_{[\bm{s}]} \, \exp\left(
    \bm{\beta_z}^\top \bm{z}_{[\bm{s}]} +
    \beta_o o_{[\bm{s}]} + \beta_w \ind_w(t) \right) +
  \gamma_0 \cdot \abs{I(\bm{s},t)} \:.
\end{equation}
The endemic model component accounts for the spatially varying population density
$\rho_{[\bm{s}]}$ and socio-economic characteristics $\bm{z}_{[\bm{s}]}$.
Additionally, we account for the quarter's distance $o_{[\bm{s}]}$ to the PUK's
quarter (1~for directly adjacent quarters, 2~for second-order neighbours, and so
forth), and include an indicator function $\ind_w(t)$ for weekends and
holidays on which fewer admissions are registered (no scheduled admissions).
By the epidemic component, each case is assumed to cause a local
increase of $\gamma_0$ in the event rate during $14$ days
in the neighbourhood of $250$ metres
around its residence.

\begin{figure}
  \centering
  \subfloat[Pearson residuals on a grid of 500m$\times$500m pixels. Positive
  values indicate an excess number of admissions compared to what is expected
  under the fitted model.\label{fig:pukfit_residuals_quadrats}]{
    \includegraphics[width=0.47\linewidth]{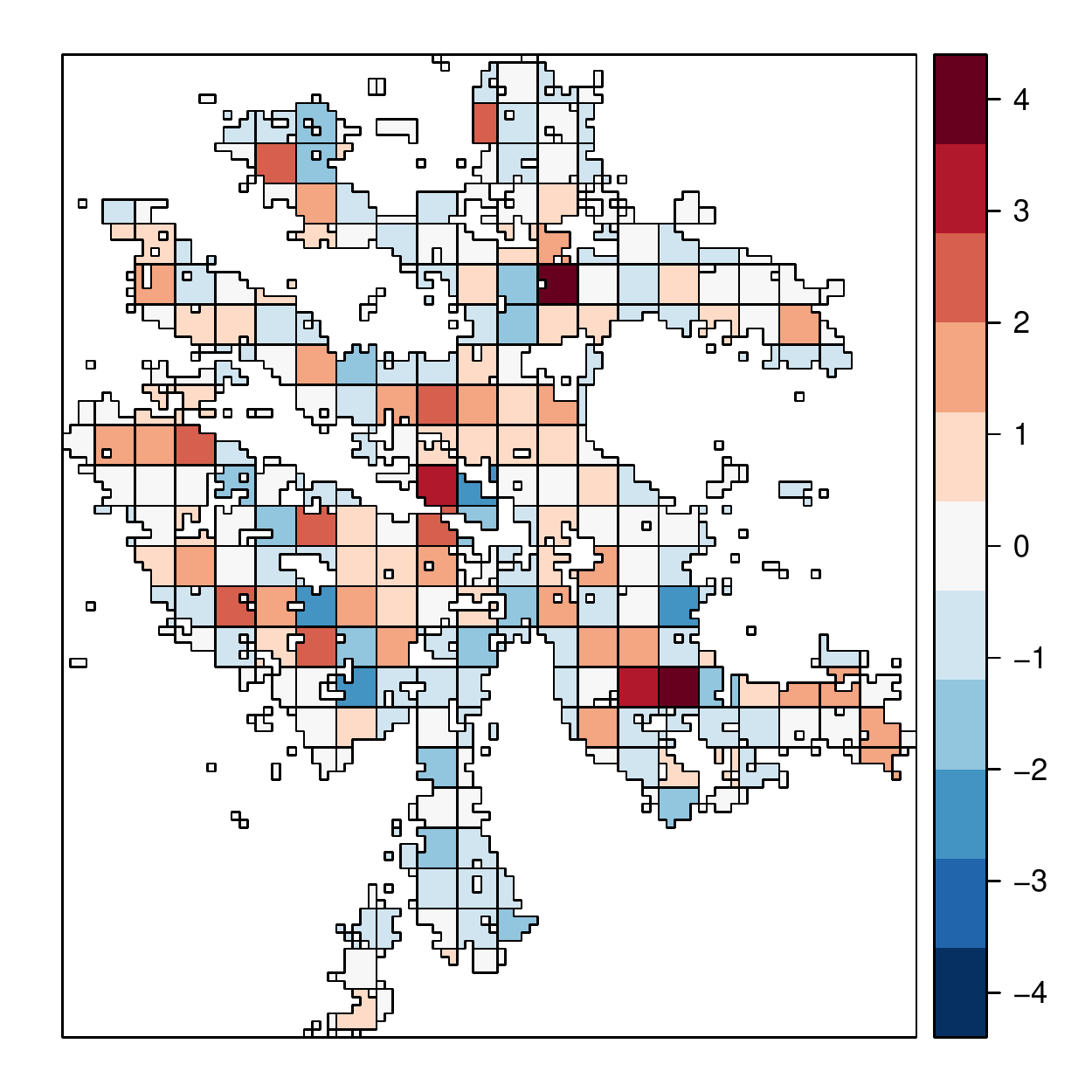}
  }
  \subfloat[Cumulative distribution of the transformed
  temporal residual process (as in \citealp{ogata1988}, Figure~10)
  under random sampling of the event times within the
  corresponding days. The dashed lines show the 95\% error bounds of the
  Kolmogorov-Smirnov statistic.\label{fig:pukfit_residuals_time}]{
    \includegraphics[width=0.47\linewidth]{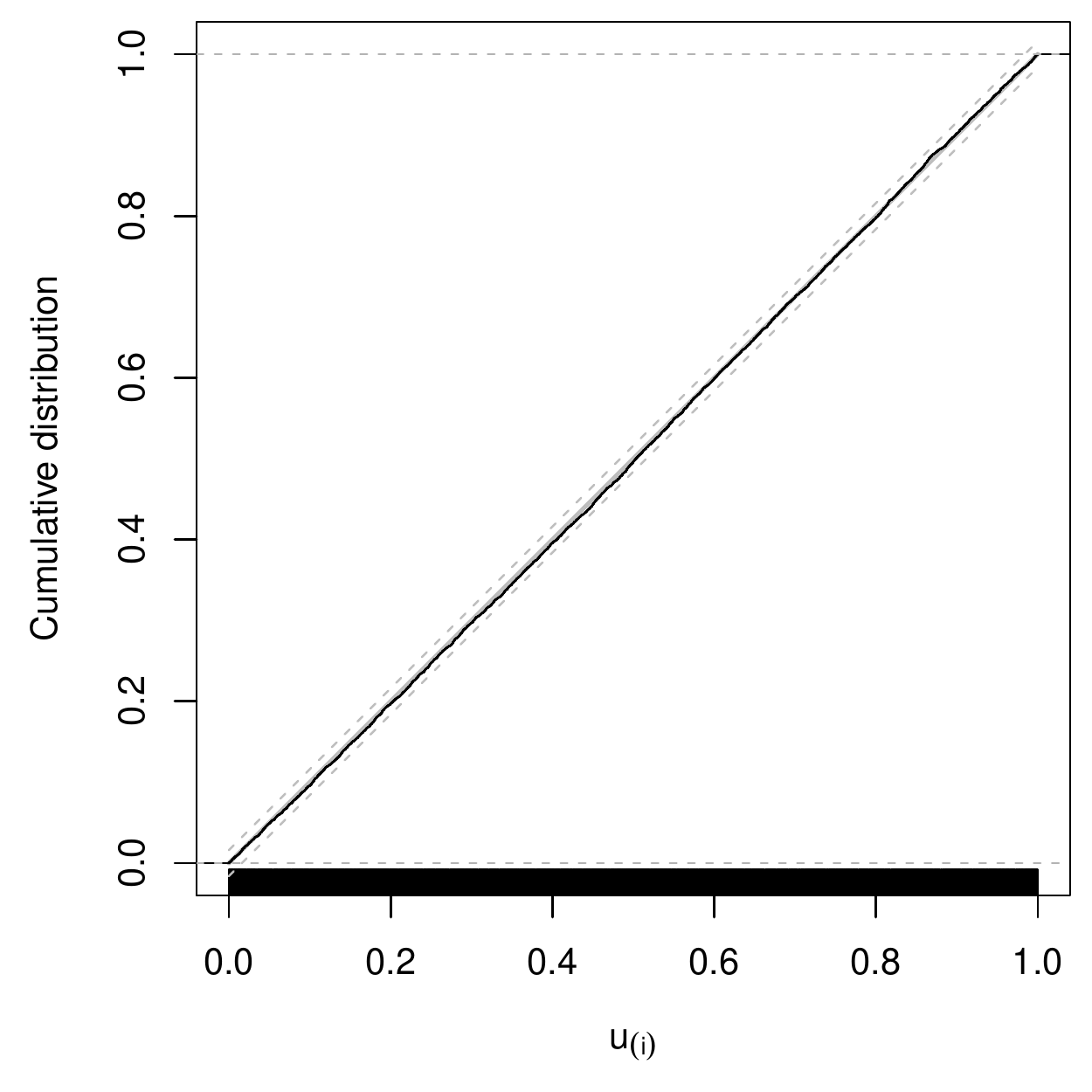}
  }
  \caption{Spatial and temporal residual analysis of the fitted model
    \eqref{eqn:pukmodel}, integrating over the other dimension.}
  \label{fig:pukfit_residuals}
\end{figure}

Fitting this model to the point pattern of all 7\,202
admissions takes only 1.3 seconds on an ordinary laptop running at 2.80GHz.
The goodness of fit can be evaluated using residual analysis methods for
space-time point processes \citep{clements.etal2011}.
Spatial pixel residuals (integrated over time) can help to identify regions with
considerably more or less cases than explained by the model
(Figure~\ref{fig:pukfit_residuals_quadrats}).
The distribution of the transformed residual process in
Figure~\ref{fig:pukfit_residuals_time} suggests that the estimated intensity
(integrated over space) provides a good description of the temporal point
pattern. 

Table~\ref{tab:pukfit_epi1} presents the rate ratio estimates for the endemic
effects of model \eqref{eqn:pukmodel}.
The event rate increases with the proportion of foreigners, inhabitants
aged 40 to 64 years, one-person households, adults employed, and people with
minimum education level. It tends to be lower in quarters with a higher mean
income and is reduced by
62\%
during weekends and holidays.
There is no evidence for further geographic
effects as reflected by the urban indicator, the distance to the PUK and the
density of psychiatric practices. 

\begin{table}[ht]
\centering
\caption{Estimated rate ratios (RR) of endemic effects.} 
\label{tab:pukfit_epi1}
\begingroup\footnotesize
\begin{tabular}{lrrr}
  \toprule
 & RR & 95\% CI & p-value \\ 
  \midrule
Urban quarter & 0.991 & 0.91 to 1.08 & 0.84 \\ 
  Distance to PUK & 0.983 & 0.96 to 1.01 & 0.16 \\ 
  \% Foreigners & 1.018 & 1.01 to 1.02 & $<$0.0001 \\ 
  \% 40--64 years & 1.084 & 1.07 to 1.10 & $<$0.0001 \\ 
  \% One-person households & 1.032 & 1.02 to 1.04 & $<$0.0001 \\ 
  Employment rate & 1.014 & 1.01 to 1.02 & 0.0011 \\ 
  \% Low-level education & 1.025 & 1.02 to 1.03 & $<$0.0001 \\ 
  Mean income (in 1000 CHF) & 0.991 & 0.98 to 1.00 & 0.021 \\ 
  Psychiatric practices / ha & 1.071 & 0.67 to 1.71 & 0.78 \\ 
  Weekend/Holiday & 0.379 & 0.35 to 0.41 & $<$0.0001 \\ 
   \bottomrule
\end{tabular}
\endgroup
\end{table}

\begin{figure}[htb]

{\centering \includegraphics[width=.9\linewidth]{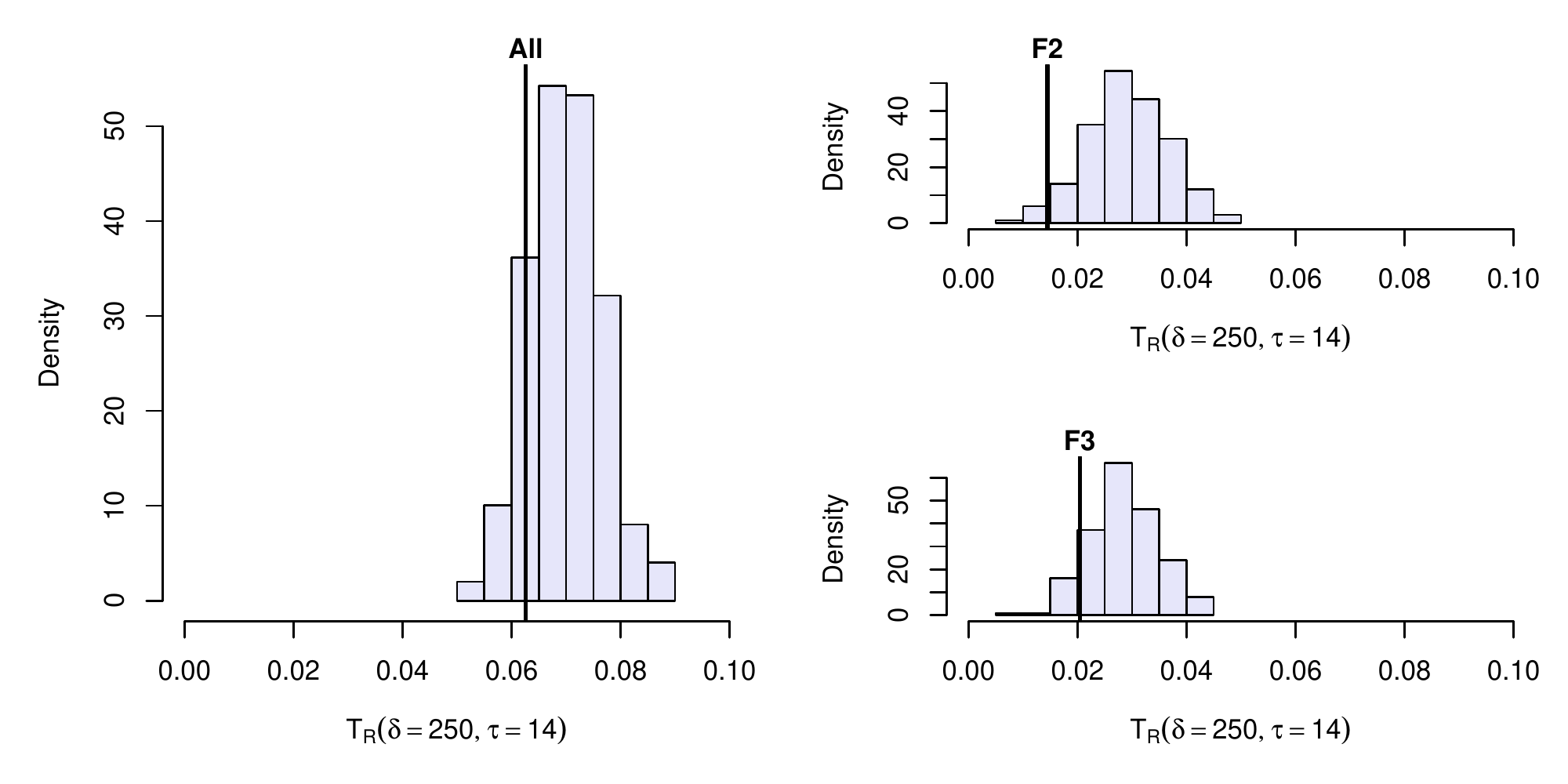} 

}

\caption{Permutation distribution of the model-based reproduction number $T_R$ (Equation~\ref{eqn:simpleR0}).}\label{fig:epitestplot}
\end{figure}

The triggering rate is estimated to be
$\hat\gamma_0 = 0.023$,
which corresponds to 
$T_R = 0.063$
secondary cases within 14 days and 250
metres.
The likelihood ratio statistic for $H_0: \gamma_0 = 0$, i.e., comparing the
endemic-only to the full model, is
$D = 115.0$.
Although the goodness of fit improves considerably with the epidemic component,
the permutation test reveals that this improvement does not reflect true
space-time interaction.
Figure~\ref{fig:epitestplot} shows the null distribution of $T_R$
from 199 data sets with permuted event times and thus no
space-time interaction by construction.
A large proportion $p=0.87$ of the permuted data sets
has a value of $T_R$ \emph{higher} than estimated for the real data.
This is in agreement with the results from the classical significance tests in
that there is no evidence for epidemicity of psychiatric hospital admissions.

Running the permutation test on the F2 and F3 subsets yields 
$p$-values of 0.97 and
0.90, respectively.
The estimated rate ratios of the endemic effects are displayed in
Table~\ref{tab:pukfit_F2F3_epi1} and are generally similar in direction and
order of magnitude as in the overall model (Table~\ref{tab:pukfit_epi1}).
There is a stronger negative association of the event rate with the distance to the
PUK, i.e., quarters further apart tend to have a lower rate of admissions
(given identical socio-economic characteristics).
For the subset of patients with affective disorders,
there is no evidence of effects of mean income,
proportion of foreigners, and employment rate on the quarter's admission rate,
which is, however, estimated to be 
19\% higher
in inner urban quarters than in outer quarters with otherwise similar characteristics.

\begin{table}[ht]
\caption{Estimated rate ratios (RR) of endemic effects in the subgroup models.}
\label{tab:pukfit_F2F3_epi1}
\centering
\footnotesize
\begin{tabular}{l|rrr|rrr}
  & \multicolumn{3}{|c|}{F2 cases} & \multicolumn{3}{|c}{F3 cases} \\
 & RR & 95\% CI & p-value & RR & 95\% CI & p-value \\ 
  \midrule
Urban quarter & 0.852 & 0.71 to 1.03 & 0.094 & 1.188 & 1.02 to 1.38 & 0.023 \\ 
  Distance to PUK & 0.947 & 0.90 to 1.00 & 0.035 & 0.927 & 0.89 to 0.97 & 0.0003 \\ 
  \% Foreigners & 1.010 & 1.00 to 1.03 & 0.17 & 1.003 & 0.99 to 1.01 & 0.61 \\ 
  \% 40--64 years & 1.140 & 1.10 to 1.18 & $<$0.0001 & 1.050 & 1.02 to 1.08 & 0.0012 \\ 
  \% One-person households & 1.049 & 1.03 to 1.07 & $<$0.0001 & 1.026 & 1.01 to 1.04 & 0.0003 \\ 
  Employment rate & 1.021 & 1.00 to 1.04 & 0.029 & 1.006 & 0.99 to 1.02 & 0.40 \\ 
  \% Low-level education & 1.035 & 1.02 to 1.05 & $<$0.0001 & 1.024 & 1.01 to 1.04 & 0.001 \\ 
  Mean income (in 1000 CHF) & 0.970 & 0.95 to 0.99 & 0.0002 & 0.996 & 0.98 to 1.01 & 0.59 \\ 
  Psychiatric practices / ha & 1.045 & 0.38 to 2.88 & 0.93 & 1.052 & 0.48 to 2.32 & 0.90 \\ 
  Weekend/Holiday & 0.505 & 0.44 to 0.58 & $<$0.0001 & 0.354 & 0.31 to 0.40 & $<$0.0001 \\ 
   \bottomrule

\end{tabular}
\end{table}

\subsection{Discussion of the results}

This observational study shows several quarter-level socio-economic characteristics to be associated
with the hospital admission rate.
Similar to previous studies suggesting that
psychiatric disorders are more frequent in socially deprived areas
\citep{reijneveld.schene1998,chaix.etal2006,sundquist.ahlen2006},
lower education and lower income were associated with an increased
hospital admission rate. However, the finding that a higher employment
rate is (weakly) associated with an increased admission rate is contradictory to
previous results \citep{kammerling.oconnor1993,simone.etal2013} and difficult to
interpret. Note that we have used the proportion of employed persons among
the population aged 15 years and over, 
whereas other studies used the unemployment rate, which is usually defined among
the labour force. For the subset of patients with affective disorders there was
no evidence for an effect of the local employment rate.
A higher proportion of foreign residents was found to increase the rate of
psychiatric admissions, which is in agreement with the recent study by
\citet{simone.etal2013}. A higher proportion of one-person households was also
found to increase the admission rate, which probably relates to a lack of social
support and/or partnership known as a protective factors.
However, previous findings on the effect of household size are contradictory
\citep{torrey.yolken1998,simone.etal2013}.

To test for space-time interaction of psychiatric admissions,
we assumed the ``spread'' to be confined to 250 metres
with homogeneous infectivity in this neighbourhood. 
This corresponds to the closest and most influential distance class in a previous
study on the spatial clustering of autism \citep{liu.etal2010}.
Regardless of whether a classical test or the model-based test was applied,
we found no indication of spatio-temporal interaction of psychiatric hospital
admissions in Zurich.
We have additionally assessed the sensitivity of the Knox and model-based tests
with respect to the assumed upper bounds of spatial and temporal interaction.
Restricting interaction to cases in the same building only ($\delta=5$ metres)
or to $\delta=50$ or 500 metres and/or assuming shorter infectious periods of
$\tau=7$ or 3 instead of 14 days led to similar test results
with no evidence for space-time interaction. For instance, the smallest
$p$-values obtained for the F2 subset are 0.28 (Knox) and 0.22 (model-based),
respectively, for the thresholds $\delta=250$ metres and $\tau=3$ days
corresponding to the highest value of $\hat{D}(\delta,\tau)$ in
Figure~\ref{fig:stKplot_PUK2}.

The spatial interaction function can be seen as a rough proxy for the
population's social contact network, which is the natural driver of
person-to-person transmission.
However, the potential interaction between subsequent admissions through social
contacts might not map well into spatially confined clusters as suggested
by the point process model. One potentially missing link is through the place
of work, where colleagues might get informed about the hospital stay.
Although we find no evidence of clustering of psychiatric admissions on the
spatio-temporal scale, interaction between cases might become
apparent if the social contact structure could be taken into account
\citep{christakis.fowler2013,coviello.etal2014}.

Another limitation of this analysis of inpatient episodes from a single hospital
is underreporting, since ``infected'' neighbours might not be
admitted to the PUK in the first place. They might turn to a general
practitioner, a registered psychiatrist, or another psychiatric institution, and
may or may not be eventually admitted to the PUK. Similar to missing links via
long-range social contacts, such unobserved cases lead to missing clusters and
underestimation of spatio-temporal interaction via the epidemic component.


To quantify the strength of interaction,
the point process model enables the estimation of an effective reproduction
number, which is also used as test statistic $T_R$. However, the value of $T_R$ 
has to be interpreted with caution since it is affected by the endemic
formulation. Specifically, the permutation test revealed that a positive force
of infection is estimated even under absence of space-time interaction.
This happens because the data do not support the piecewise constant endemic intensity
on the chosen grid. Although we have restricted the observation region
to the populated area of Zurich, the population density actually varies
within each quarter. Given the large amount of 7\,202 cases,
there is evidence that subsequent cases tend to be closer to a previous case
than implied by a spatially homogeneous intensity simply due to spatial
clustering of the population.
The permutation test accounts for this endemic misspecification 
in the null distribution of the test statistic
by keeping the marginal spatial and temporal locations fixed.
If significant space-time interaction is found (as for the occurrence of
invasive meningococcal disease, see Supplementary material),
the difference between the observed $T_R$
and its average value under permutation could be used as an effect size
quantifying true space-time interaction.
Note that for data conforming to the piecewise constant endemic intensity, for
example, for simulations from the fitted endemic-epidemic point process model,
the null distribution of $T_R$ is centred around zero (see Supplementary material).

\section{Conclusions}
\label{sec:conclusions}

Classical significance tests of space-time clustering solely
operate on the events' spatial and temporal distance matrices to evaluate
the association of spatial and temporal closeness.
In this paper we have demonstrated how spatio-temporal interaction can be
investigated more thoroughly using a two-component point process regression model.
While its endemic component explains the background risk of new events through
regional and/or time-varying covariates, the superposed epidemic component
reflects that events tend to appear in spatio-temporal clusters.
Hence, the model-based global clustering test is adjusted for, e.g.,
socio-economic heterogeneity and seasonal effects.

For the psychiatric hospital admissions in Zurich, we found several
quarter-level socio-economic characteristics to be associated with the
endemic occurrence of cases. However, there was no evidence for
spatio-temporal interaction of the cases.
The hypothesized social influence on help-seeking behaviour is thus not apparent
in local clusters of sequential hospital admissions.
Therefore, standard methods for non-contagious spatial processes could be used
to analyse these data, e.g., spatial point pattern analysis,
or disease mapping and ecological regression approaches
\citep{Waller.Gotway2004,Baddeley.etal2015}.
A simple example is a Poisson regression model for the number of patients by quarter using the
population as a multiplicative offset and the socio-economic characteristics
as explanatory variables (see, e.g., \citealp[Section~7.7]{lawson2013},
\citealp{marshall1991}, or \citealp{wakefield2007}).
In mathematical terms, the expected
number of patients (the Poisson rate) of quarter $i$ could be modelled as
$\lambda_i = n_i \cdot \exp(\bm{\beta}^\top \bm{z}_i)$, where $n_i$ is the
quarter's population and $\bm{z}_i$ the vector of socio-economic
characteristics. Doing so, the estimated risk ratios and confidence intervals
are in fact very similar to those from the endemic-epidemic 
point process model presented in Table~\ref{tab:pukfit_epi1}.
Alternatively, 
a separable \emph{endemic-only} point process model could also serve as input for further
exploratory analysis of second-order properties based on the inhomogeneous
$K$-function of \citet{gabriel.diggle2009}, which requires an estimate of
the intensity $\lambda(\bm{s}_i,t_i)$ at all events.

If no individual cases but only area-level counts over several time periods
are available, the tests for spatial, temporal and spatio-temporal clustering
proposed by \citet{raubertas1988} could be used.
The areal time-series model of
\citet[Section~3]{meyer.held2013} also offers means of quantifying the importance of
transmission from neighbouring regions and estimating the spatial distance decay
of interaction. 

The methods discussed in this paper are "general tests of clustering" not to be
confused with "tests for the detection of clusters" \citep{besag.newell1991}.
\citet{kulldorff2006} reviews both of these for purely spatial processes.
Methods to detect emergent space-time clusters have been proposed by, e.g.,
\citet{piroutek.etal2014} monitoring the cumulative sum of a local Knox
statistic, 
or \citet{kulldorff.etal2005} using a space-time scan statistic.

Although the presented endemic-epidemic point process regression framework
is not intended to replace the convenient classical tests,
its added value is manifold:
First, it offers insight into endemic characteristics of the process via
modelling the background rate using covariates for which the test is adjusted.
Second, a distance decay of interaction can be estimated and incorporated
into the test. An example is the power-law kernel that we have used to reflect
large-scale human travel in modelling the occurrence of invasive meningococcal
disease in Germany \citep{meyer.held2013}. For comparison, we have also applied
the various test procedures to these data and find clear evidence of
spatio-temporal interaction as expected for an infectious disease (see
Supplementary material).
Last but not least, if there is evidence for space-time interaction,
the model allows for a more detailed estimation of event-specific
infectiousness through the epidemic predictor $\eta_j$.
The local population density or event characteristics, such as the patient's age
group or disease severity, could be associated with the triggering rate, such
that events are expected to produce a varying number of secondary cases.
Whether the model-based test also has more power to detect space-time
interaction than the classical tests -- especially if ``mechanistic'' knowledge
about the process and suitable covariates enable a well-fitting model --
is an open question to be possibly answered by future simulation studies.




\section*{Acknowledgements}

We thank Michael Höhle (Stockholm University) and two anonymous referees for
illuminating comments on a previous version of this manuscript,
as well as Dominik Ullmann (Swiss Federal Statistical Office)
and Michael Böniger (City of Zurich Statistics Office)
for providing the socio-economic data of Zurich's quarters.
The work of the first author was financially supported by
the Swiss National Science Foundation (Project \#137919).



\appendix
\section{Software}
\label{sec:software}

All analyses were performed using the statistical software environment
\proglang{R} \citep{R:base}.
We have implemented the model-based test
(\code{epitest}) as well as the Knox test (\code{knox}) in the framework of the
\proglang{R} package \CRANpkg{surveillance} \citep{R:surveillance}.
The package contains methods for visualization,
inference and simulation of the endemic-epidemic point process model
\citep[as described in][Section~3]{meyer.etal2014},
as well as a wrapper (\code{stKtest}) to perform $K$-function
analysis using the \CRANpkg{splancs} package \citep{R:splancs}.
For all these tests, our implementations allow the $B$ permutations to be
distributed across multiple cores.
To perform the standardized Mantel test, we used a \proglang{C} implementation
(\code{mantel.randtest}) from the package \CRANpkg{ade4} \citep{chessel.etal2004} 
as at version 1.7-4.

Embedding the test for interaction in a point process regression model for the
conditional intensity function comes at the cost of a considerably increased
runtime compared to the classical tests (Table~\ref{tab:runtimes}).
Fitting a single model (with constant spatial interaction function) only takes
about a second for a given data set, but the permutation approach requires the
two competing models (with and without the epidemic component) to be
re-estimated on each of the $B$ permutations.
Furthermore, the extra long runtime for the whole data set (also for the Knox
test) is due to memory-intensive distance matrix calculations, which do not
scale well on a laptop. This will, however, very much improve by a more
efficient implementation of the permutation handling planned for future versions
of the \proglang{R} package \pkg{surveillance}.

\begin{table}[htb]
  \centering
  \caption{Runtime comparison of the various tests for space-time interaction.
    The timings refer to the real elapsed time in minutes.
    Note that a \proglang{C} implementation was used for the Mantel test; for
    the other tests, the $B$ permutations were distributed across four cores
    ($B=999$ for the classical tests and $B=199$ for the
    model-based approach).}
  \label{tab:runtimes}
  \footnotesize
  \begin{tabular}{r|rrrr}
    \toprule
    & Knox & Mantel & $K$-function & Model-based\\
    \midrule
    All cases ($n=7\,202$)   & 20.6 & 1.7 & 8.8 & 44.0\\
    F2 cases ($n=1\,465$) & 0.6 & 0.1 & 0.3 & 1.6\\
    F3 cases ($n=2\,180$) & 1.6 & 0.2 & 0.7 & 3.0\\
    \bottomrule
  \end{tabular}
\end{table}

Shapefiles have been edited using \proglang{QGIS} 2.2 \citep{QGIS} with the
\proglang{fTools} plugin 0.6.2. The original administrative boundaries
have been simplified using \href{www.MapShaper.org}{MapShaper.org} v.\ 0.1.18
\citep{harrower.bloch2006} to speed up computations.
Within \proglang{R}, the packages
\CRANpkg{sp} 
\citep{R:sp},
\CRANpkg{rgdal} 
\citep{R:rgdal},
\CRANpkg{rgeos} 
\citep{R:rgeos},
and \CRANpkg{spatstat} 
\citep{Baddeley.etal2015}
were used to deal with geographic shapes.
The list of holidays in Zurich was obtained from package
\CRANpkg{timeDate} \citep{R:timeDate}.
Tables have been created with the \CRANpkg{xtable} package \citep{R:xtable}.
This manuscript has been generated dynamically in R version 3.2.3 (2015-12-10)
using \CRANpkg{knitr} 
\citep{knitrbook}.



\section*{Supplementary material}

A supplementary application of the various tests for space-time interaction
to cases of invasive meningococcal disease,
can be found in the publisher's version of record at
\href{http://dx.doi.org/10.1016/j.sste.2016.03.002}{doi:10.1016/j.sste.2016.03.002}.



\renewcommand{\bibfont}{\small}
\bibliography{references,R}

\end{document}